\documentclass[12pt]{article}
\usepackage{amsmath}
\usepackage{amssymb}
\usepackage{amsthm}
\usepackage{amsbsy}
\usepackage{amsfonts}
\newtheorem{theorem}{Theorem}

\newtheorem{lemma}[theorem]{Lemma}
\newtheorem{conjecture}[theorem]{Conjecture}

\begin{document}

\title{On Darboux integrable semi-discrete systems of exponential type }
\author{ Kostyantyn Zheltukhin$^1$ and Ergun Bilen$^2$\\
\small $^1$ Department of Mathematics, Faculty of Science,\\
\small Middle East Technical University 06531 Ankara, Turkey.\\
\small e-mail: zheltukh@metu.edu.tr\\
\small $^2$  Department of Mathematics, Faculty of Science,\\
\small Middle East Technical University 06531 Ankara, Turkey.\\
\small e-mail: ergun.bilen@metu.edu.tr}

\begin{titlepage}

\maketitle

\begin{abstract}
In the present paper we consider a discretization of hyperbolic systems of exponential type. We proved that, in the case of $2\times 2$ systems, 
the resulting semi-discrete system is Darboux integrable only if it corresponds to a Cartan matrix of a semi-simple Lie algebra.  
\end{abstract}


\medskip


\end{titlepage}

\section{Introduction}

In a recent paper \cite{HZhY} Habibullin et all proposed a method for discretization of hyperbolic type exponential systems
\begin{equation}\label{cont_system}
r^i_{xy}=e^{\sum a_{ij}r^j}.
\end{equation}
It was conjectured that the method preserves integrability of such systems. The integrability of system \eqref{cont_system} and its discretization is understood as Darboux integrability \cite{Dar}.
It was shown in \cite{ShYa} that system \eqref{cont_system} is Darboux integrable if and only if $A$ is a Cartan matrix of a semi-simple Lie algebra.
To check for Darboux integrability we use the notion of characteristic rings introduced in \cite{ShYa}, \cite{LSSh}. Later the notion of characteristic ring was applied to classify
Darboux integrable hyperbolic equations in \cite{SokZh}-\cite{KZh} , see also a review paper \cite{ZhMHSh}. For discrete equations the Darboux integrability  was considered in \cite{Adler}, \cite{HP}. 
The notion of characteristic rings for discrete systems and its application to classification of such systems was developed in \cite{H}-\cite{HZh}      

We note that the discrete analogs of system \eqref{cont_system} find applications in many areas of physics,  see a review paper \cite{KNS}.

Let us give necessary definitions.
Consider an $N\times N $ matrix $A=(a_{ij})$ and a corresponding  exponential system.
Following \cite{HZhY} we decompose $A$ into a sum $A=A_+ + A_-$,
where $A_+=(a^+_{ij})$ is an upper triangular matrix with all diagonal entries equal to $1$ and $A_-=(a^-_{ij})$ is a lower triangular matrix. The corresponding differential-difference exponential type system is
\begin{equation}\label{gen_dis_system}
q^i_{1x}-q^i_x=e^{\sum a^+_{ij}q_1^j+\sum a^-_{ij}q^j}.
\end{equation}
In this equation $q^i(n,x)$, $i=1,\dots N$, are unknown functions depending on discrete variable $n$ and continuous variable $x$, and $q^i_{1}=q^i(n+1,x)$ denotes a shift in $n$ and
$q^i_{x}=\frac {d}{dx}q^i(n,x)$ denotes derivative with respect to $x$.
\begin{conjecture} (see \cite{HZhY})
A system \eqref{gen_dis_system} is Darboux integrable if and only if $A$ is a Cartan matrix of a semi-simple Lie algebra.
\end{conjecture}

As was shown in \cite{HZhY}  if  $N=2$ and $A$ is a Cartan matrix of a semi-simple Lie algebra then the corresponding system \eqref{gen_dis_system} is Darboux integrable.
In the present paper we prove that for $N=2$ Darboux integrability of the system \eqref{gen_dis_system} implies that $A$ is a Cartan matrix of a semi-simple Lie algebra.
Our proof is based on the notion of the characteristic $x$-ring. We recall that a hyperbolic system is Darboux integrable if and only if its characteristic $x$- and $n$-rings are finite dimensional (see \cite{HP}). It turns out that it is enough to consider only the characteristic $x$-ring. The characteristic $x$-ring is finite dimensional if and only if $A$  is a Cartan matrix of a semi-simple Lie algebra.
\begin{theorem}  \label{maintheorem}
When $N=2$, the system \eqref{gen_dis_system} is Darboux integrable if and only if the matrix $A = (a_{ij})_{2x2}$ is the Cartan matrix.
\end{theorem}

The paper is organized as follows in Section 2 we describe the characteristic $x$-ring corresponding to the system \eqref{gen_dis_system} with general matrix $A$, with  $N=2$, and in Section 3
we determine for which matrices $A$ the characteristic $x$-ring is finite dimensional and we prove the above theorem.

\section{Characteristic $x$-ring}.

Let $ A = (a_{ij})$, $i,j=1,2$, be an arbitrary matrix.
Consider a system
\begin{equation}\label{4006}
\begin{aligned}
u_{1x} - u_{x} &= e^{(a_{11}-1)u + u_{1} + a_{12}v_{1}} \\
v_{1x} - v_{x} &= e^{a_{21}u + (a_{22}-1)v + v_{1}}.\\
\end{aligned}
\end{equation}
We assume that $a_{12},a_{21} \neq 0$, otherwise the system reduces to a pair of ordinary differential equations.
In what follow we use notation $u_k=u(n+k,x)$, $v_k=v(n+k,x)$ and $u_{kx}=\frac{d}{dx} u_k$,  $v_{kx}=\frac{d}{dx} v_k$.

The characteristic $x$-ring for system \eqref{4006}, denoted $\cal X$, is generated by the  vector field (see  \cite{HP})
\begin{equation*}
Z = \frac{\partial}{\partial x} + \sum_{k=-\infty}^{\infty}(u_{kx}\frac{\partial}{\partial u_{k}} +v_{kx}\frac{\partial}{\partial v_{k}}), \label{4007}
\end{equation*}
which corresponds to the total derivative with respect to $x$, and vector fields
\begin{equation*}
Y_{1} = \frac{\partial}{\partial u_{x}}, \quad\quad Y_{2} = \frac{\partial}{\partial v_{x}}. \label{4010}
\end{equation*}
It follows from  \eqref{4006} that
\begin{align*}
u_{kx} & = u_{x}  +\sum_{i =1}^{k} e^{(a_{11}-1)u_{i-1} + u_{i} + a_{12}v_{i}}, &  u_{-kx} &= u_{x}  - \sum_{i =1}^{k} e^{(a_{11}-1)u_{-i} + u_{-i+1} + a_{12}v_{-i+1}} \\
v_{kx} &= v_{x}  +\sum_{i =1}^{k} e^{(a_{22}-1)v_{i-1} + v_{i} + a_{21}u_{i-1}}, &   v_{-kx} &= v_{x}  -\sum_{i =1}^{k} e^{(a_{22} -1)v_{-i} + v_{-i+1} + a_{21}u_{-i}} .
\end{align*}
For convenience we define
\begin{align*}
 M_{i} &= e^{(a_{11}-1)u_{i-1} + u_{i} + a_{12}v_{i}}, & M_{-i} &= e^{(a_{11}-1)u_{-i} + u_{-i+1} + a_{12}v_{-i+1}} \\
 N_{i} &= e^{(a_{22}-1)v_{i-1} + v_{i} + a_{21}u_{i-1}}, & N_{-i} &= e^{(a_{22} -1)v_{-i} + v_{-i+1} + a_{21}u_{-i}},
\end{align*}
$i=1,2 \dots $.
 So the vector field $Z$ takes form
\begin{multline}
Z =\frac{\partial}{\partial x} +u_x\frac{\partial}{\partial u} + \sum_{k=1}^{\infty}\left( u_{x}  +  \sum_{i=1}^{k} M_i \right)\frac{\partial}{\partial u_{k}} +  
 \sum_{k=1}^{\infty}\left( u_{x}  - \sum_{i =1}^{k} M_{-i} \right)\frac{\partial}{\partial u_{-k}}  \\
 + v_x\frac{\partial}{\partial v} +\sum_{k=1}^{\infty}\left( v_{x}  +\sum_{i =1}^{k} N_i \right)\frac{\partial}{\partial v_{k}} +
 \sum_{k=1}^{\infty}\left(v_{x}  -\sum_{i =1}^{k} N_{-i} \right)\frac{\partial}{\partial v_{-k}}.
\end{multline}

Taking commutator of $Z$ with $Y_1$ and $Y_2$ we obtain vector fields
\begin{equation*}
A = [Y_{1}, Z] = \sum_{j=-\infty}^{\infty}\frac{\partial}{\partial u_{j}},\qquad B = [Y_{2}, Z] = \sum_{j=-\infty}^{\infty}\frac{\partial}{\partial v_{j}}
\end{equation*}
The commutators of $Z$ with $A$ and $B$ give us vector fields
\begin{multline*}
[A,Z] = a_{11}\sum_{k=1}^{\infty}\left( \sum_{i =1}^{k} M_i \right)\frac{\partial}{\partial u_{k}}  -
a_{11}\sum_{k=1}^{\infty}\left( \sum_{i =1}^{k}M_{-i} \right)\frac{\partial}{\partial u_{-k}}  \\
+ a_{21}\sum_{k=1}^{\infty}\left(\sum_{i =1}^{k}N_i\right)\frac{\partial}{\partial v_{k}} -
a_{21}\sum_{k=1}^{\infty}\left(\sum_{i =1}^{k}N_{-i} \right)\frac{\partial}{\partial v_{-k}}
\end{multline*}
and
\begin{multline*}
[B,Z] = a_{12}\sum_{k=1}^{\infty}\left( \sum_{i =1}^{k} M_i \right)\frac{\partial}{\partial u_{k}}  -
a_{12}\sum_{k=1}^{\infty}\left( \sum_{i =1}^{k} M_{-i} \right)\frac{\partial}{\partial u_{-k}}  \\
+ a_{22}\sum_{k=1}^{\infty}\left(\sum_{i =1}^{k} N_i \right)\frac{\partial}{\partial v_{k}} -
a_{22}\sum_{k=1}^{\infty}\left(\sum_{i =1}^{k} N_{-i} \right)\frac{\partial}{\partial v_{-k}}.
\end{multline*}

It is convenient to replace vector fields $[A,Z]$ and $[B,Z]$ by their linear combinations
\begin{equation*}
P_{1} = \frac{1}{a_{22}a_{11} - a_{21}a_{12}}(a_{22}[A,Z] - a_{21}[B,Z])
\end{equation*}
and
\begin{equation*}
P_{2} = \frac{1}{a_{22}a_{11} - a_{21}a_{12}}(-a_{12}[A,Z] + a_{11}[B,Z]).
\end{equation*}
So, vector  fields $P_{1}$ and $P_{2}$ are
\begin{equation*}
P_{1} = \sum_{k=1}^{\infty}\left( \sum_{i =1}^{k} M_i \right)\frac{\partial}{\partial u_{k}}  -
\sum_{k=1}^{\infty}\left( \sum_{i =1}^{k} M_{-i} \right)\frac{\partial}{\partial u_{-k}}
\end{equation*}
and
\begin{equation*}
P_{2} =\sum_{k=1}^{\infty}\left( \sum_{i =1}^{k} N_i \right)\frac{\partial}{\partial v_{k}} -
\sum_{k=1}^{\infty}\left(\sum_{i =1}^{k} N_{-i} \right)\frac{\partial}{\partial v_{-k}} \, .
\end{equation*}
Since $Z=\frac{\partial}{\partial x} + u_xA + v_xB + P_1 + P_2$,
we can consider vector fields $\frac{\partial}{\partial x}$, $Y_{1}$, $Y_{2}$, $A$, $B$, $P_{1}$ and $P_{2}$ as generators of the characteristic $x$-ring $\cal X$.

The vector fields $\displaystyle{\frac{\partial}{\partial x}}$, $Y_{1}$, $Y_{2}$, $A$ and $B$ are pairwise commuting and vector fields $\frac{\partial}{\partial x}$, $Y_{1}$, $Y_{2}$ commute with
vector fields $P_{1}$, $P_{2}$.
The commutators  of $A$, $B$ with $P_{1}$, $P_{2}$ are
\begin{equation*}
[A,P_1]=a_{11}P_1,\quad [B,P_1]=a_{12}P_1,\quad [A,P_2]=a_{21}P_2,\quad [B,P_2]=a_{22}P_2\, .
\end{equation*}
Thus the characteristic $x$-ring $\cal X$ is finite dimensional if and only if the ring generated by vector fields $P_1$, $P_2$, denoted $\cal A$, is finite dimensional.

If we define a sequence of linear subspaces
\begin{equation*}
\mathbb{X}_{0} = Lin \bigl\{P_{1}, P_{2}\bigl\},
\end{equation*}
and
\begin{equation*}
\mathbb{X}_{n} = Lin \bigl\{ [P_{i}, V] : V\in \mathbb{X}_{n-1},\, i=1,2 \bigl\},  \quad  n=1,2\dots \, .
\end{equation*}
then it is easy to see that
\begin{equation}\label{A-decomposition}
\mathcal{A} = \mathbb{X}_{0} \oplus \mathbb{X}_{1} \oplus \mathbb{X}_{2} + ... \quad .
\end{equation}
We define sequences of vector fields $T_n$, $R_n$, $n=1,2, \dots $, by
\begin{equation*}
T_n=[P_1,T_{n-1}], \, n=2,3\dots \quad \mbox{and} \quad T_{1} = [P_{1}, P_{2}],
\end{equation*}
\begin{equation*}
R_n=[P_2,R_{n-1}], \, n=2,3 \dots \quad \mbox{and} \quad R_{1} = T_1
\end{equation*}

\section{Proof of Theorem \ref{maintheorem}}

From the decomposition \eqref{A-decomposition} we see that $\cal A$ is finite dimensional if and only if $\mathbb X_n=0$ for some $n$.
The following lemma (see \cite{HP}) is very useful to derive conditions  arising from $\mathbb X_n=0$.

\begin{lemma} \label{lemma1}
Suppose that the vector field
\begin{equation*}
K = \sum_{k=1}^{\infty} \Bigl(\alpha_{k}\frac{\partial}{\partial u_{k}} + \alpha_{-k}\frac{\partial}{\partial u_{-k}}\Bigl) + \sum_{k=1}^{\infty} \Bigl(\beta_{k}\frac{\partial}{\partial v_{k}} + \beta_{-k}\frac{\partial}{\partial v_{-k}}\Bigl)
\end{equation*}
satisfies the equality $DKD^{-1} = hK$, where $h$ is a function depending on shifts and derivatives of variables $u$ and $v$, then $K = 0$.
\end{lemma}

To apply this lemma we evaluate how the transformation $D(*)D^{-1}$ acts on  vector fields.
For vector fields  $A$, $B$, $P_{1}$ and $P_{2}$ we have
\begin{equation}  \label{40006}
 DAD^{-1} = A, \quad DBD^{-1} = B
\end{equation}
and
\begin{equation}  \label{400061}
 DP_{1}D^{-1} = P_{1} - M_{1}A, \quad DP_{2}D^{-1} = P_{2} - N_{1}B
\end{equation}
Using \eqref{40006} and \eqref{400061} we calculate action of  the transformation $D(*)D^{-1}$ on the vector fields $T_n$ and $R_n$.

First we consider vector fields $T_n$, $n=1,2,\dots\, .$ We find
\begin{multline}\label{4016aa}
DT_{1}D^{-1}= D[P_{1}, P_{2}]D^{-1} = [DP_{1}D^{-1}, DP_{2}D^{-1}] = [P_{1} - M_{1}A  ,P_{2} - N_{1}B]  \\
 = T_{1} +a_{12}N_{1}P_{1} - a_{21}M_{1}P_{2} + a_{21}M_{1}N_{1}B.
\end{multline}
In the same way as for $T_1$ we find
\begin{multline}\label{DT2D-1}
DT_{2}D^{-1} =  T_{2} - (2a_{21} +a_{11})M_1T_{1}- a_{12}(a_{11} + 2a_{21})P_{1} +a_{21}(a_{11} + a_{21} -1)M_1^2P_2+\\ a_{21}(1-a_{11}-a_{21})M_1^2N_1B
\end{multline}
For other vector fields $T_n$ we have following lemma.

\begin{lemma}
For the vector fields $T_n$, n=3,4\dots, we have
\begin{equation}\label{DTnD-1}
DT_{n}D^{-1} = T_{n} + \alpha_nM_1T_{n-1} + \beta_nM_{1}^{2}T_{n-2} + ... \, ,
\end{equation}
where
\begin{equation}\label{alpha-n}
\alpha_n= -\frac{n(n-1)}{2}a_{11} - na_{21},
\end{equation}
\begin{multline}\label{beta-n}
\beta_n = \frac{1}{24}n\bigg(-8a_{11} + 12na_{11} - 4n^{2}a_{11} - 2a_{11}^{2} + 9na_{11}^{2} - 10n^{2}a_{11}^{2} + 3n^{3}a_{11}^{2} \\
+ 12a_{21} - 12na_{21} + 12a_{11}a_{21} - 24na_{11}a_{21}+12n^{2}a_{11}a_{21} - 12a_{21}^{2} + 12na_{21}^{2}\bigg).
\end{multline}
\end{lemma}

\noindent{\bf Proof.}
For $n=3$ we have
\begin{multline}
DT_{3}D^{-1}=  [DP_{1}D^{-1}, DT_{2}D^{-1}] = \\
[P_{1} - M_{1}A, T_{2} - (2a_{21} +a_{11})T_{1}- a_{12}(a_{11} + 2a_{21})P_{1} +a_{21}(a_{11} + a_{21} -1)P_2+ \dots]  \\
 = T_{3} +\alpha_3T_2 +  \beta_3T_1 +\dots
\end{multline}
where $\alpha_3=-3(a_{11}+a_{21})$ and $\beta_3=\left(a_{11}(2a_{11} + 6a_{21} -1) + a_{21}(3a_{21}-3)\right)$. Thus,  $\alpha_3$ satisfies \eqref{alpha-n},
 $\beta_3$ satisfies \eqref{beta-n}.

For $n\ge 4$,  we can  show, using induction,  that the coefficients $ \alpha_n$ and $\beta_n$ satisfy  recursion relations
\begin{eqnarray}
\alpha_{n} &=& \alpha_{n-1} - (na_{11} + a_{21}),   \\
\beta_{n} &=& \beta_{n-1} + (1-na_{11} - a_{21})\alpha_{n-1} 
\end{eqnarray}
Solving the above  recursion relations  we obtain that $\alpha_n$ satisfies \eqref{alpha-n} and
 $\beta_n$ satisfies \eqref{beta-n}. $\Box$\\

Now we consider vector fields $R_n$, $n=1,2,\dots\,$. We have
\begin{equation}
DR_{1}D^{-1}= DT_1D^{-1}= R_{1} +a_{12}N_{1}P_{1} - a_{21}M_{1}P_{2} + a_{21}M_{1}N_{1}B.
\end{equation}
In the same way as in \eqref{4016aa} we find
\begin{multline} \label{DR2D-1}
DR_{2}D^{-1} = R_2-(2a_{12}+a_{22})N_1R_1+a_{12}(1-a_{22}-a_{12})N_1^2P_2
\end{multline}
and
\begin{multline}\label{DR3D-1}
DR_{3}D^{-1} = R_{3}  - (3a_{12} + 3a_{22})N_{1}R_2 + (a_{12} + 2a_{22} -1)(2a_{12} + a_{22})N_{1}^{2}R_1  \\ + a_{12}(2 - 3a_{22})(1 - a_{12} - a_{22})N_{1}^{3}P_{2} + a_{12}(1 - a_{12} - a_{22})N_{1}^{4}B
\end{multline}
For other vector fields $R_n$ we have the following lemma.

\begin{lemma}
For the vector fields $R_n$, n=4,5 \dots, we have
\begin{equation}\label{DRnD-1}
DR_{n}D^{-1} = R_{n} + \tilde \alpha_nN_1R_{n-1} + \tilde\beta_nN_{1}^{2}R_{n-2} +\tilde\gamma_nN_{1}^{3}R_{n-3} + ... \, ,
\end{equation}
where
\begin{equation}\label{til-alpha-n}
\tilde\alpha_n= -\frac{n(n-1)}{2}a_{22} - na_{12},
\end{equation}
\begin{multline}\label{til-beta-n}
\tilde \beta_n = \frac{1}{24}(-2+n) \left( 3 n^3 a_{22}^2-4 n^2 a_{22} (1+a_{22}-3 a_{12})+12 a_{12} (-1+a_{22}+a_{12})\right.\\
+ \left. n (4 a_{22}+a_{22}^2+12 (-1+a_{12}) a_{12})\right)
\end{multline}
\begin{multline}\label{til-gamma-n}
\tilde \gamma_n = -\frac{1}{48} (-3+n) (-2+n) (n^4 a_{22}^3-2 n^3 a_{22}^2 (2+a_{22}-3 a_{12})    \\
 + 8 a_{12} (4-5 a_{22}+a_{22}^2-6 a_{12}+3 a_{22} a_{12}+2 a_{12}^2)   \\
+ n^2 a_{22} (4+6 a_{22}+a_{22}^2-20 a_{12}-2 a_{22} a_{12}+12 a_{12}^2)  \\
+ 2n(a_{22}^2 (-1+6 a_{12})+4 a_{12} (2-3 a_{12}+a_{12}^2)+a_{22} (-2-6 a_{12}+6 a_{12}^2))).
\end{multline}
\end{lemma}

\noindent{\bf Proof.}
By direct calculations for $n=4$ we have
\begin{equation}
DR_4D^{-1} =R_{4} +\tilde\alpha_4 R_{3} + \tilde\beta_4 R_{2} + \tilde\gamma_4 R_{1}+ \dots
\end{equation}
where $\tilde\alpha_4=-(6a_{22}+4a_{12})$, $\tilde\beta_4=(11a_{22}^{2} +5(-1+a_{12})a_{12}+a_{22}(-4+17a_{12}))$ and
$\tilde\gamma_4=-(-1 + 2 a_{22} + a_{12}) (-2 + 3 a_{22} + a_{12}) (a_{22} + 2 a_{12})$.
It is easy to check that $\tilde\alpha_4$, $\tilde\beta_4$ and $\tilde\gamma_4$ satisfy equalities \eqref{til-alpha-n}, \eqref{til-beta-n} and \eqref{til-gamma-n}.
For $n\ge 5$  we can  show, using induction,  that the coefficients $ \tilde\alpha_n$, $\tilde\beta_n$ and $\tilde\gamma_n$ satisfy  recursion relations
\begin{eqnarray}
\tilde \alpha_{n+1} &=& \tilde\alpha_n - (na_{22} + a_{12}),  \\
\tilde\beta_{n+1} &=& \tilde\beta_n + (1-na_{22}-a_{12})\tilde\alpha_n, \\
\tilde\gamma_{n+1} &=& \tilde\gamma_{n} + (2-na_{22}-a_{12})\tilde\beta_n. 
\end{eqnarray}
Solving these recursion relations we obtain  equalities \eqref{til-alpha-n}, \eqref{til-beta-n} and \eqref{til-gamma-n}.  $\Box$\\

The previous two lemmas allow us to specify matrix $A=(a_{ij})$ in the case of finite characteristic $x$-ring.

\begin{lemma}\label{lemma12}
Let $\mathcal A$ be a finite dimensional ring then for some integer $n\ge 2$  we have $T_n=0$ and $T_{n-1}\ne 0$. In that case
\begin{align}
 a_{11} & = 2,  & a_{12} & = -1,  \\
 a_{21} &= n-2, & a_{22} & = 2.
\end{align}
\end{lemma}

\noindent {\bf Proof.}
First we show that $a_{11}=2$ and $a_{21}=n-2$.
If $n=2$ we have $T_2=0$ and  $DT_2D^{-1}=0$.
So coefficients of $T_1$ and $P_2$ in \eqref{DT2D-1} must be zero. Thus
\begin{eqnarray}
2a_{11}+a_{21}&=&0,\\
a_{11}+a_{21}-1&=&0.
\end{eqnarray}
Solving these equalities for  $a_{11}$ and $a_{12}$ we get  $a_{11}=2$ and $a_{21}=n-2$.
In the same way for $n>2$, we have $T_n=0$ and $DT_nD^{-1}=0$.
Hence coefficients $\alpha_n$ and $\beta_n$ in the equality \eqref{DTnD-1} must be zero. Using equalities \eqref{alpha-n} and \eqref{beta-n} we have
\begin{equation*}
 -\frac{n(n-1)}{2}a_{11} - na_{21}=0,
\end{equation*}
\begin{multline}
\beta_n = \frac{1}{24}n\bigg(-8a_{11} + 12na_{11} - 4n^{2}a_{11} - 2a_{11}^{2} + 9na_{11}^{2} - 10n^{2}a_{11}^{2} + 3n^{3}a_{11}^{2}\\
 + 12a_{21} - 12na_{21} + 12a_{11}a_{21} - 24na_{11}a_{21}+12n^{2}a_{11}a_{21} - 12a_{21}^{2} + 12na_{21}^{2}\bigg)=0.
\end{multline}
Solving these equalities for  $a_{11}$ and $a_{21}$ we get  $a_{11}=2$ and $a_{21}=n-2$.

Now let us show that $a_{22}=2$ and $a_{12}=-1$. Since $\mathcal A$ is finite dimensional there exists $m\ge2$ such that $R_m=0$. It turns out that $m=2$.

Indeed, assume $m>2$. If $m=3$ then we have $R_3=0$. We have $DR_3D^{-1}=0$.
So coefficients of $R_1$, $R_2$ and $P_2$ in \eqref{DR3D-1} must be zero. Equating the coefficients to zero we get overdetermined system
\begin{eqnarray*}
3a_{12} + 3a_{22}&=&0,\\
(a_{12} + 2a_{22} -1)(2a_{12} + a_{22})&=&0,\\
(2 - 3a_{22})(1 - a_{12} - a_{22})&=&0
\end{eqnarray*}
that has no solution. Thus it is not possible to have $m=3$.

In the same way if $m>3$, we have $R_m=0$ and $DR_mD^{-1}=0$.
Hence coefficients $\tilde \alpha_m$, $\tilde\beta_m$ and $\tilde\gamma_m$ in the equality \eqref{DRnD-1} must be zero. Using equalities
\eqref{til-alpha-n}, \eqref{til-beta-n} and \eqref{til-gamma-n} for $\tilde \alpha_m$, $\tilde\beta_m$ and $\tilde\gamma_m$ respectively we get an overdetermined system
\begin{equation}
\tilde \alpha_m=0,\quad \tilde\beta_m=0, \quad \tilde\gamma_m=0
\end{equation}
that has no solution. Thus it is not possible to have $m>3$.

If $m=2$ then we have $R_2=0$ and $DR_2D^{-1}=0$. So coefficients of $R_1$ and $P_2$ in \eqref{DR2D-1} must be zero. Thus
\begin{eqnarray}
2a_{12}+a_{22}&=&0,\\
1-a_{22}-a_{12}&=&0.
\end{eqnarray}
Solving these equalities for  $a_{11}$ and $a_{12}$ we get  $a_{11}=2$ and $a_{21}=n-2$. $\Box$\\ 

 In the proof of previous lemma we see that in case of finite dimensional characteristic $x$-ring all vector fields $R_{n}=0$, $n\geq2$. The similar statement is true for vector fields $T_{n}, n\geq4$. To show that we have the following lemma.

\begin{lemma}\label{lemma13}
Let $\mathcal A$ be a finite dimensional ring. Then $T_{n}=0$, $n\geq4$.
\end{lemma}

\noindent {\bf Proof.}
Let us introduce sequences 
\begin{equation*}
W_{n} = [P_{2}, W_{n-1}], \quad n= 4,5 \dots \quad \mbox{with} \quad W_{3} = T_{3}, 
\end{equation*}
and
\begin{equation*}
V_{n} = [P_{2}, V_{n-1}], \quad n= 5,6 \dots \quad \mbox{with} \quad V_{4} = T_{4}.
\end{equation*}
First we show that $W_{n} = 0$, for $n\geq5$. We have  
\begin{equation*}
[P_2,T_2]=[P_2,[P_1,T_1]]=-([T_1,[P_2,P_1]]+ [P_1,[T_1,P_2]])=[T_1,T_1]+[P_1,R_2]=0,
\end{equation*}
since $R_2=0$ in the case of the finite dimensional ring $\mathcal A$. Taking commutator of $P_1$ with $[P_2,T_2]$ we get
\begin{equation*}
[P_1,[P_2,T_2]]=-([T_2,[P_1,P_2]]+[P_2,[T_2,P_1]])=0,
\end{equation*}
 which gives $W_4=-[T_1,T_2]$. So, for $W_5$ we have
\begin{multline*} 
 W_{5} = [P_{2}, -[T_1,T_2]] = [T_1, [T_2, P_2]] + [T_2, [P_2, T_1]]= -[T_{1},[P_2,T_2] ] + [T_{2},R_{2}] = 0.
\end{multline*} 
This shows that $W_5=0$ and  $W_n=0$ for $n>5$. 

Now we show that the assumption $T_4\ne 0$ leads to a contradiction. Assume $T_4\ne 0$. Since $\mathcal A$ is finite dimensional we have 
$V_n=0$ for some $n\ge 5$. Assume $V_5=0$. Then considering   
\begin{equation*}
DV_{5}D^{-1} = V_{5} -(4a_{12} + a_{22})N_{1}T_{4} + \dots \,  ,
\end{equation*}
we have $(4a_{12} + a_{22})=0$ and by Lemma \ref {lemma12} we have  $a_{12}=-1$ and $a_{22}=2$, which is not possible. 
Assume $V_n=0$, for $n>5$. Then evaluating $DV_nD^{-1}$, using induction, we get 
\begin{equation*}
DV_{n}D^{-1}= V_{n} + \frac{1}{2}[(42 - 8n)a_{22} + (7n - n^{2} -22)a_{12}] N_{1}V_{n-1} +  \dots \, .
\end{equation*}  
 We have $[(42 - 8n)a_{22} + (7n - n^{2} -22)a_{12}]=0$ and by Lemma \ref {lemma12} we have $a_{12}=-1$ and $a_{22}=2$, which is not possible.  $\Box$ \\

\noindent {\bf Proof of  Theorem \ref{maintheorem}.} Since $\mathcal A$ is finite dimensional, using Lemma \ref {lemma12} and Lemma \ref {lemma13} we find that
\begin{equation}
\begin{aligned}
 a_{11} & = 2,  & a_{12} & = -1,  \\
 a_{21} &= -c, & a_{22} & = 2,
\end{aligned} 
\end{equation}
where $ c=1,2,3$.
For each value of $c$, $A$ is the Cartan matrix and the corresponding $x-$ and $n-$ integrals are described in \cite{HZhY}. Hence that completes the proof of the main theorem.

\end{document}